\PassOptionsToPackage{final}{graphicx}

\documentclass[10pt, conference]{IEEEtran}
\IEEEoverridecommandlockouts
\usepackage{etex}

\usepackage{amsfonts}
\usepackage{amsmath}
\usepackage{amssymb}
\usepackage{booktabs}
\usepackage[color]{changebar}
\usepackage{color} % For the \warn command.
\usepackage[outline]{contour}
\usepackage{environ}
\usepackage{etoolbox}
\usepackage{flushend}
\usepackage{fmtcount}
\usepackage{graphicx}
\usepackage{latexsym}
\usepackage[final]{listings}
\usepackage[final]{microtype}
\usepackage{multicol}
\usepackage{newtxtext,newtxmath}
\usepackage[all]{nowidow}
\usepackage{outlines}
\usepackage{pgfplots}
\usepackage{pgfplotstable}
\usepackage[binary-units]{siunitx}
\usepackage{stmaryrd}
\usepackage{tabulary}
\usepackage{tikz}
\usepackage{tikz-qtree}
\usepackage{todonotes}
\usepackage{totcount}
\usepackage{url}
\usepackage[inline]{enumitem}
\setlist{nosep,leftmargin=*}

\usepackage[final, pdfusetitle]{hyperref}

\usepackage{changebar-float-fix}
\usepackage{ieee-natbib}
\usepackage{ieee-subcaption}
\usepackage{refs}

\usepackage{cleveref}

\usepackage{hypernat}
\usepackage{textcomp}

\cbcolor{red}

\newcommand{\Omit}[1]{}

\newcommand{\OnlySubmission}[1]{}

\newcommand{\OnlySupplementary}[1]{}

\newcommand{\warn}[1]{\textbf{\color{red} [#1] }}

\newcommand{\name}[0]{Source Forager} % chktex 20

\newcommand{\Ed}[2]{\ensuremath{\text{ed}(#1, #2)}}
\newcommand{\Pre}[1]{\ensuremath{\text{pre}(#1)}}
\newcommand{\Post}[1]{\ensuremath{\text{post}(#1)}}
\newcommand{\Size}[1]{\ensuremath{\text{size}(#1)}}

\newcommand{\algo}[0]{\texttt{algo-qs}}
\newcommand{\libc}[0]{\texttt{libc-qs}}

\newcommand{\configuration}[1]{\emph{#1}}
\newcommand{\equalAll}[0]{\configuration{equal-all}}
\newcommand{\dynSelect}[0]{\configuration{dyn-select}}
\newcommand{\randSelect}[0]{\configuration{rand-select}}
\newcommand{\svmWeights}[0]{\configuration{svm-weights}}
\newcommand{\svmWeightsCross}[0]{\configuration{cross-weights}}
\newcommand{\soloC}[0]{\configuration{solo-$c$}}
\newcommand{\dynNonl}[0]{\configuration{dyn-nonl}}
\newcommand{\svmNonl}[0]{\configuration{svm-nonl}}

\newsavebox{\savedlistingbox}
\newcommand{\listingbox}[1]{%
  \savebox{\savedlistingbox}{%
    \footnotesize
    \lstinputlisting{#1}}%
  \usebox{\savedlistingbox}}

\definecolor{finding}{gray}{.9}
\newlength{\findingskip}
\newlength{\findingtextwidth}
\NewEnviron{finding}[1]{%
  \vspace{2\findingskip}
  \noindent
  \fcolorbox{black}{finding}{%
    \parbox{\findingtextwidth}{%
      \textbf{\Cref{#1} Finding:} \BODY}}
  \vspace{\findingskip}}

\Crefformat{enumi}{\textbf{#2#1#3}}
\crefformat{enumi}{\textbf{#2#1#3}}
\crefformat{section}{\textsection#2#1#3}
\crefname{figure}{Fig.}{Figs.}
\crefname{table}{Table}{Tables}
\crefname{equation}{Eq.}{Eqs.}

\usetikzlibrary{
  arrows,
  backgrounds,
  calc,
  decorations.pathreplacing,
  fit,
  positioning,
  scopes,
  shapes.geometric,
  shapes.multipart,
  shapes.symbols,
}

\contourlength{1pt}

\pgfplotsset{
  compat=1.9,
  retrieval plot/.style={
    every node near coord/.append style={
      /pgf/number format/fixed,
      /pgf/number format/fixed zerofill,
      /pgf/number format/precision=2,
      anchor=east,
      font=\scriptsize,
      rotate=90,
    },
    haloed nodes near coords/.style={
      nodes near coords=\contour{white}{\pgfmathprintnumber{\pgfplotspointmeta}},
    },
    legend style={cells={anchor=west}},
    x tick label style={rotate=90},
    xtick=data,
    ylabel=MAP score,
    ymajorgrids,
    ymax=1,
    ymin=0,
  },
}

\pgfplotstableset{
  col sep=comma,
  columns/Selection Rate/.style=percentage column,
  ignore chars={',\%},
  every first column/.style=text column,
  every head row/.style={
    before row=\toprule,
    after row=\midrule,
  },
  every last row/.style={
    after row=\bottomrule,
  },
  inline tabular data/.style={
    col sep=&,
    row sep=\\,
  },
  multicolumn names,
  percentage column/.default=2,
  percentage column/.style={
    column type={S[table-format=#1,table-space-text-post=\%]},
    multiply by=100,
    numeric as string type,
    postproc cell content/.append style={
      /pgfplots/table/@cell content/.add={}{\%},
    },
  },
  text column/.style={
    column type=l,
    multicolumn names=l,
  },
  string type,
}

\hypersetup{
  pdfauthor={Vineeth Kashyap, David Bingham Brown, Ben Liblit, David Melski, and Thomas Reps},
  pdfkeywords={code search, similar code, program features},
  pdfsubject={mining, big data, and recommendation systems; search-based and knowledge-based software engineering
},
}

\begin{document}

\title{\name{}: A Search\\Engine for Similar Source Code
  \texorpdfstring{%
    \thanks{%
      Supported, in part, by a gift from Rajiv and Ritu Batra; by AFRL
      under DARPA MUSE award FA8750-14-2-0270, % chktex 8
      and by the UW--Madison Office of the Vice Chancellor for
      Research and Graduate Education with funding from the Wisconsin
      Alumni Research Foundation.  Any opinions, findings, and
      conclusions or recommendations expressed in this publication are
      those of the authors, and do not necessarily reflect the views
      of the sponsoring agencies.  T.\ Reps has an ownership interest
      in GrammaTech, Inc., which has licensed elements of the
      technology reported in this publication.}}{}}

\author{
\IEEEauthorblockN{Vineeth Kashyap\IEEEauthorrefmark{1},
David Bingham Brown\IEEEauthorrefmark{2},
Ben Liblit\IEEEauthorrefmark{2}, 
David Melski\IEEEauthorrefmark{1}, and
Thomas Reps\IEEEauthorrefmark{1}\IEEEauthorrefmark{2}}
\IEEEauthorblockA{\IEEEauthorrefmark{1}GrammaTech, Inc., Ithaca, New York, USA\\ 
Email: \{vkashyap,melski\}@grammatech.com}
\IEEEauthorblockA{\IEEEauthorrefmark{2}University of Wisconsin--Madison, USA\\
Email: \{bingham,liblit,reps\}@cs.wisc.edu}
}

\maketitle

%!TEX root = paper.tex
\begin{abstract}
Developers spend a significant amount of time searching for
code---e.g., to understand how to complete, correct, or adapt their own code for a new context.
Unfortunately, the state of the art in code search has not evolved
much beyond text search over tokenized source.
Code has much richer structure and semantics than normal text,
and this property can be exploited to specialize the code-search
process for better querying, searching, and ranking of code-search
results.

We present a new code-search engine named \name{}.
Given a query in the form of a C/C++ function, \name{} searches a pre-populated code database for similar C/C++ functions.
\name{} preprocesses the database to extract a variety of simple
code features that capture different aspects of code.
A search returns the $k$ functions in the database that
are most similar to the query, based on the various extracted code features.

We tested the usefulness of \name{} using a variety of
code-search queries from two domains.
Our experiments show that the ranked results returned by \name{} are accurate,
and that query-relevant functions can be reliably retrieved even when searching
through a large code database that contains very few query-relevant
functions.

We believe that \name{} is a first step towards much-needed
tools that provide a better code-search experience.
\end{abstract}

%%% Local Variables: 
%%% mode: latex
%%% End: 

%  LocalWords:  pre 's


\IEEEpeerreviewmaketitle{}

%!TEX root = paper.tex
\section{Introduction}
\label{Se:Introduction}

In this age of software proliferation, it is useful to be able to
search large source-code corpora effectively for code with desired
properties.\footnote{
  In this paper, the term ``search'' is used
  in the sense of Google search---namely, to retrieve documents
  that are related to a specified query.
  ``Search'' is \emph{not} used in the sense of finding an occurrence
  of a user-specified string or pattern in a given document.
}
Developers routinely use code search as a learning and
debugging tool for tasks such as looking for existing functionality in
a code base, determining how to use an API or library, gathering
information about what code is intended to do, etc.~\citep{FSE:SSE2015}.

Text-based search techniques are not always precise enough for code
because they focus purely on strings in the code: comments, complete
or partial names of functions and variables, and so on.  Text search
largely ignores code structure and semantics (i.e., what the code does
and how it does it).  A text-based approach can cause searching to be
imprecise:  relevant code fragments may be missed, while many spurious
matches may be returned.  Recent search techniques allow users to
specify certain aspects of code semantics in addition to the textual
query~\cite{DMKD:LBN+2009,ICSE:R2009,SEM:SEB2014,OOPSLA:SC2006,HCIIR:B2007,ASE:LBO+2007,ICSE:HM2005}.
Some techniques allow users to specify structural requirements, such
as that the search target should have nested loops.  Others specify
context, such as that the search target should implement a particular
interface.  Yet others specify sets of input/output pairs.

Additional semantic information can improve search accuracy. 
However, existing techniques share the following shortcomings:
\begin{itemize}
\item The techniques do not provide a unified way of specifying
  semantics for the search query.  Each technique has its own ad-hoc
  specification of the semantic aspects of the code that it uses.

\item Each technique is closely married to its chosen semantic aspect,
  which is deeply ingrained into the implementation of the search
  technique.  This tight coupling makes it hard to extend these
  techniques to model additional semantic aspects.
\end{itemize}

We propose a search technique for finding similar source code that addresses these shortcomings:
\begin{itemize}
\item \textbf{Unified Query Specification.}  Our code-search mechanism
  takes code fragments as queries. Various kinds of semantic
  information can be extracted from the query and used by the search.
  This approach provides a unified mechanism for code search:
  \emph{searching code using code fragments}.  Moreover, the same
  techniques for extracting semantic information are used on both
  queries and elements of the corpus being searched, leading to
  greater consistency.

\item \textbf{Extensibility.}  Our code-search technique uses a vector
  of \emph{feature-observations} extracted from elements in the
  corpus.  Feature-observations capture various aspects of the syntax
  and semantics of a program (each such aspect is called a
  feature-class), and provide a unified interface for querying.  This
  approach also makes our search technique extensible:  it is easy to
  introduce more feature-classes that model additional aspects of the
  code.
\end{itemize}

In addition to being useful on its own right as a developer tool, similar-code search can serve as an important building block for automated program repair and program synthesis.
The ability to find other code similar to a query can help automated tools learn from the similar code, and fix bugs or perform code completion tasks on the query.  

The main contributions of \name{} are:
\begin{itemize}
\item The ability to perform C/C++ code searches using code fragments
  as queries. The searches and answers of \name{} are both based on a query
  formalism that is close to the concepts that developers are already
  familiar with.  \Omit{The queries thus have a smaller cognitive
    distance to the search task and the results.}

  \item
    A code-search architecture that uses multiple code feature-classes simultaneously.
    The architecture is extensible, allowing easy addition of new code feature-classes, 
    which enhances the dimensions along which code is searched.
  
  \item
    A mechanism for automatically selecting useful code feature-classes to be employed in code search of a given query, given no a priori domain information about the query.

  \item A supervised-learning technique to pre-compute the relative
  importance of different feature-classes, when it is known that a
  query belongs to a specific domain for which suitable training data
  is available.
\end{itemize}

\Omit{
One of the techniques used in \name{} makes use of transfer learning
\cite{TLDD:PY10}:
\name's algorithm for \emph{code search} can make use of weights
derived from parameters obtained from a Support Vector Machine
(SVM) that was trained to identify \emph{code clusters}.    
}

\regtotcounter{section}
\newcounter{remaining}
\defcounter{remaining}{\totvalue{section} - 1} % chktex 8
\newcommand{\remaining}[0]{\ifnumless{\value{remaining}}{0}{\textbf{??}}{\numberstring{remaining}}}

\paragraph*{Organization}
The remainder of the paper is organized into \remaining{} \namecrefs{Se:RelatedWork}:
\sectref{Overview} gives an overview of our approach and algorithms.
\sectref{CodeSearch} describes the methods in detail.
\sectref{Experiments} presents our experimental results.
\sectref{RelatedWork} discusses related work.
\Omit{
\sectref{Conclusion} concludes.
}

%!TEX root = paper.tex
\section{Overview}
\label{Se:Overview}

\name{} is a search engine for finding similar source code.
It takes an input query as C/C++ source text, then
searches a pre-populated database for similar C/C++ code, returning a
ranked list of results.
The units of code about which \name{} can reason about are called
\emph{program elements}.
In its current incarnation, program elements are C/C++ functions;
that is, both queries and results are C/C++ functions.

\begin{figure}
  \resizebox{\linewidth}{!}{\newcommand{\vstrut}[0]{\vphantom{f}}

\begin{tikzpicture}[
  >=stealth,
  ellipsis/.style={
    font=\bfseries,
    rotate=90,
  },
  every label/.style={
    align=flush center,
    font=\scriptsize,
  },
  feature vector/.style={
    draw,
    rectangle split,
    rectangle split parts=7,
    rectangle split horizontal,
    rounded corners=2,
  },
  rotated stage/.style={
    anchor=north,
    rotate=90,
    stage,
  },
  stage/.style={
    draw,
    rounded corners=4,
    align=flush center,
  },
  ]

  % two feature vectors in database, with "..." between them
  \node (upper corpus vector) [feature vector] {} ;
  \node (lower corpus vector) [feature vector, below=of upper corpus vector] {} ;
  \node (corpus vector ellipsis) [at=($(upper corpus vector)!.5!(lower corpus vector)$), ellipsis] {\dots} ;

  % database cylinder surrounding two feature vectors
  \node (database) [
  aspect=.2,
  cylinder,
  draw,
  fit=(upper corpus vector) (lower corpus vector),
  minimum height=80,
  minimum width=80,
  shape border rotate=90,
  ] {} ;

  % label atop database
  \node [at=($(database.before top)!.5!(database.after top)$)] {code database} ;

  % weight determination stage, below database
  \node (weight determination) [below=.3 of database, stage] {feature-class weight \\ determination} ;
  \draw (database.200) edge [->, bend right=45] (weight determination.west) ;

  % query vector, below weight determination
  \node (query vector) [
  below=.7 of weight determination,
  feature vector,
  label={below:feature-observations of \\ various feature-classes},
  ] {} ;

  % stretched horizontal brace above query vector
  \draw [decorate, decoration={amplitude=10, brace}]
  ($(query vector.north west)!.4!(weight determination.south west)$) --
  ($(query vector.north east)!.4!(weight determination.south east)$) ;

  % anchor points for arrows to elements of various feature vectors
  \node (upper corpus program) [below left=.1 and 2 of upper corpus vector] {} ;
  \node (lower corpus program) [above left=.1 and 2 of lower corpus vector] {} ;
  \node (query program) [above left=.1 and 2 of query vector] {} ;

  % neighbor search stage, right of weight determination
  \node (neighbor search) [right=of weight determination, rotated stage] {similarity-based neighbor search} ;
  \draw [->] (database.340) -- (database.340 -| neighbor search.north) ;
  \draw [->] (weight determination) -- node [above, font=\scriptsize] {weights} (neighbor search) ;
  \draw [->] (query vector) -- (query vector -| neighbor search.north) ;

  % results stage, right of neighbor search
  \node (results) [right=.5 of neighbor search.south, rotated stage] {similar-code results} ;
  \draw [->] (neighbor search) -- (results) ;

  % feature extraction engine stage, left of lower edge of database
  \node (feature extraction engine) [fill=white, fill opacity=.7, left=1.5 of database.south west, minimum width=150, rotated stage, text opacity=1] {feature extraction engine} ;

  % left anchor points for arrow combs into corpus vectors
  \path (0, -.2) + (feature extraction engine.north |- upper corpus vector.south) coordinate (upper corpus vector comb) ;
  \path (0, +.2) + (feature extraction engine.north |- lower corpus vector.north) coordinate (lower corpus vector comb) ;
  \path (0, +.2) + (feature extraction engine.north |- query vector.north) coordinate (query vector comb) ;

  % corpus cloud, left of feature extraction engine
  \node (corpus) [at=(corpus vector ellipsis), cloud, cloud ignores aspect, draw, fill=white, xshift=-130] {corpus\vstrut} ;

  % arrows from corpus cloud to feature extraction engine
  { [->, on background layer]
    \draw [->] (corpus |- upper corpus vector comb) -- (upper corpus vector comb) ;
    \draw [->] (corpus |- lower corpus vector comb) -- (lower corpus vector comb) ;
  }
  \node [at=({$(corpus.east)!.5!(feature extraction engine.north)$} |- corpus vector ellipsis), ellipsis, label={right:program \\ elements}] {\dots} ;

  % query stage, below corpus cloud
  \node (query) [at=(corpus |- query vector comb), stage] {query\vstrut} ;
  \draw [->] (query) -- (query vector comb) ;

  % arrow combs to elements of various feature vectors
  { [->, on background layer]
    \foreach \slot in {0,...,6} {
      \draw (upper corpus vector comb) -| ($(upper corpus vector.one south)!\slot/6!(upper corpus vector.seven south)$) ;
      \draw (lower corpus vector comb) -| ($(lower corpus vector.one north)!\slot/6!(lower corpus vector.seven north)$) ;
      \draw (query vector comb) -| ($(query vector.one north)!\slot/6!(query vector.seven north)$) ;
    }
  }
\end{tikzpicture}}
  \caption{Overview of the \name{} architecture\label{fig:overview}}
\end{figure}

\Cref{fig:overview} provides an architectural overview of \name{}.
\name{} has two stages:  an offline phase to populate its code
database, and an online query-search phase.

\subsection{Offline Phase: Population of the \name{} Database}
\label{Se:offline}

In this phase, \name{} analyzes a given code corpus, and populates a
code database with rich information about each of the functions in the
code corpus.  \name{} extracts several different kinds of information
about each function; we refer to each of the different kinds of
information as a \emph{feature-class}.  \Cref{Se:CodeSearch} describes
our different feature-classes in detail.  A \emph{feature-observation}
is some specific value observed for a given feature-class.  Thus, each
function has one feature-observation for each feature-class.  For
example, one of our feature-classes is Numeric Literals.  The
corresponding feature-observation is the set of all the numeric
constants used in the function.  For the binary-search implementation
code given in \Cref{fig:code-example}, the Numeric Literals
feature-observation is the set $\{-1, 0, 1, 2\}$.

\begin{figure}[tb]
  \centering
  \listingbox{binsearch.c}
  \caption{Example program that implements a binary search over a sorted integer array\label{fig:code-example}}
\end{figure}

A feature extraction engine consists of several \emph{feature extractors},
which collect a given function's
feature-observations into a \emph{feature-vector}.
Note that the elements of the feature-vector can be non-numeric,
such as sets, multisets, trees, maps, etc.
The number of feature-classes determines the length of the feature-vector.

The feature extractors operate on a code corpus, and populate a code
database.  Each element of the code database consists of a C/C++
function from the corpus along with its extracted feature-vector.  If
Numeric Literals is employed as one of the feature-classes, then one
element of a function's feature-vector is
the set of numeric constants.

The code database also has access to several \emph{similarity
  functions}, one for each feature-class.  The similarity function for
a given feature-class takes any two feature-observations belonging to
that feature-class and returns a value between $0.0$ and $1.0$.  A
higher value indicates greater similarity between two
feature-observations.  For example, the similarity function for
Numeric Literals is the Jaccard index.  Given two sets $S_1$ and
$S_2$, the Jaccard index is given by:
\begin{equation} \label{eq:jaccard}
{sim}_{\emph{Jacc}(S_1, S_2)} = \frac{|S_1 \cap S_2|}{{|S_1 \cup S_2|}}.
\end{equation}

\subsection{Online Phase: Search for Similar Code}
\label{Se:online-search}

In the online search phase, \name{} takes a query and uses the same
feature-extraction infrastructure to obtain the feature-vector that
corresponds to the query.  This infrastructure reuse creates a
consistent representation and view of code throughout the code-search
infrastructure.  For each feature-class in the feature-vector, a
weight is assigned to determine the importance of that feature-class.
This feature-class weight determination is based on which
configuration \name{} is run with;
\cref{subsec:configurations,Se:PerQuerySelection,Se:svm-weights}
provide an overview of the different configurations.

A combined similarity function is defined on any two feature-vectors
by combining the per-feature-class similarity functions with the per-feature-class weight assignment using a weighted average.
That is,
\begin{equation} \label{eq:combined-sim}
  sim_{\emph{combined}}(\vec{A}, \vec{B})
  =
  \frac{\sum_{c=1}^{n_{\textit{cl}}} sim_{c}(\vec{A}_c, \vec{B}_c) \cdot w_c}{\sum_{c=1}^{n_{\textit{cl}}} w_c},
\end{equation}
where $\vec{A}$ and $\vec{B}$ are two feature-vectors;
$n_{\textit{cl}}$ is the total number of feature-classes (i.e, the length of each feature-vector);
$sim_{c}$ is the similarity function for feature-class $c$;
$\vec{A}_c$ and $\vec{B}_c$ are the feature-observations
for feature-class $c$ in $\vec{A}$ and $\vec{B}$, respectively;
and $w_c$ is the weight assigned to feature-class $c$.

The feature-vector of the query is compared with each of the
feature-vectors in the code database using this combined similarity
function, and the $k$ most-similar functions (that is, with the
highest similarity scores to the query) are returned as results (for
some configurable limit $k$).  \cref{fig:example-result} shows an
example \name{} code-search result when the code in
\cref{fig:code-example} is used as query.

\begin{figure}[tb]
  \centering
  \listingbox{binsearch_result.c}
  \caption{Example \name{} code-search result for the query in
    \cref{fig:code-example}.  This result is a recursive
    implementation of binary search.\label{fig:example-result}}
\end{figure}

We have two implementations of \name{}.  The first one is a
slower-performing version, in which the code database is implemented
as a large, in-memory JSON~\citep{JSON} object, and the various
similarity functions and the algorithm for $k$-most-similar
function-search are implemented in Python.  This implementation allows
for easier and quicker experimentation with new ideas.  We use this
version for the experiments reported in \cref{Se:Experiments}.

The second implementation integrates our infrastructure with
Pliny-DB~\cite{PDB}, which is an in-memory object-store database
implemented in C++.  The feature-observations in feature-vectors are
serialized into efficient in-memory data structures by Pliny-DB\@.
Pliny-DB has access to similarity functions implemented in C++ for all
feature-classes.  It implements the search for the $k$ functions most
similar to the query by
\begin{enumerate*}[label=(\arabic*)] % chktex 36
\item scanning all the feature-vectors in the database,
\item comparing each of them to the query feature-vector, and
\item maintaining a priority queue of size $k$ that keeps track of the
  $k$ most-similar feature-vectors.\end{enumerate*}
Given a query feature-vector and relative weights for different
feature-classes, Pliny-DB can find the 10 most-similar functions in a
code database containing \num{500000} functions in under 2 seconds on
a single machine with 8 Intel i7 \SI{3.6}{\giga\hertz} cores and
\SI{16}{\giga\byte} RAM\@.  Effort is underway by the developers of
Pliny-DB to make a distributed version, which would allow \name{} to
search large code databases without taking a big performance hit: a
large code database can be split into $p$ smaller units that can each
be searched in parallel, and the sorted $k$ most-similar results from
each of the $p$ units can be merged using a multi-way merge algorithm.

\subsection{Extensible Architecture}

\name{}'s architecture allows for easy extension.  To add a new
feature-class, one implements (1) a feature extractor that determines
the feature-observation for any given function, and (2) a
corresponding similarity function.  We currently implement our feature
extractors using CodeSonar\textsuperscript{\textregistered}.  However,
\name{} is not tightly coupled with CodeSonar: any C/C++ processing
tool can be used to implement a feature extractor.  The
feature-observations for all existing feature-classes are represented
with well-known container data structures, such as lists, maps, and
trees; all similarity functions work at the level of container data
structures, and thus are available to be reused with any additional
user-supplied feature extractors.  Furthermore, \name{} is not tied to
having functions as the only kind of program element.  The underlying
architecture is also not limited to C/C++, and thus \name{} can be
re-targeted to perform code searches of programs written in other
languages.

%!TEX root = paper.tex
\section{Code Search}
\label{Se:CodeSearch}

In this section, we first describe the different feature-classes and
the accompanying similarity functions that are employed in \name{}.
We then describe two configurations of \name{}.  The first
configuration (\dynSelect{}) selects a subset of the feature-classes on
a per-query basis for performing code search: this configuration is
useful when no additional information is available regarding a code
query.  The second configuration (\svmWeights{}) pre-computes the
relative importance of feature-classes for a specific domain ahead of
time using supervised-learning techniques.  This configuration is useful when the
domain of the code query is known.

\subsection{Feature-Classes and Similarity Functions}
\label{Se:FeatureClassesAndSimilarityFunctions}

\Cref{tab:feature-classes} summarizes \name's feature-classes.  Below,
we further describe these feature-classes and their associated
similarity functions.

\begin{table}
  \centering
  \caption{A brief overview of the different feature-classes employed in \name{}. The marked* feature-classes all use Jaccard index (\cref{eq:jaccard}) as the similarity function. The similarity functions used for the remaining feature-classes accompany their descriptions in~\cref{Se:FeatureClassesAndSimilarityFunctions}.\label{tab:feature-classes}}
  \pgfplotstabletypeset [
  after row=\addlinespace,
  begin table=\begin{tabulary}{\linewidth}, end table=\end{tabulary},
  columns/Feature-Class/.style=wrapped column,
  columns/Brief Description/.style=wrapped column,
  inline tabular data,
  wrapped column/.style={column type=L, multicolumn names=l},
  ] {
    Feature-Class & Brief Description \\
    Type--Operation Coupling* & types used and operations performed on the types \\
    Skeleton Tree & structure of loops and conditionals \\
    Decorated Skeleton Tree & structure of loops, conditionals, and operations \\
    Weighted NL Terms & processed natural language terms in code \\
    3 Graph CFG BFS & CFG subgraphs of size 3, BFS used for generating subgraphs \\
    4 Graph CFG BFS & CFG subgraphs of size 4, BFS used for generating subgraphs \\
    3 Graph CFG DFS & CFG subgraphs of size 3, DFS used for generating subgraphs \\
    4 Graph CFG DFS & CFG subgraphs of size 4, DFS used for generating subgraphs \\
    Modeled Library Calls* & calls made to modeled libraries \\
    Unmodeled Library Calls* & calls made to unmodeled libraries \\
    User-Defined Library Calls* & calls made to user-defined libraries \\
    Type Signature & input types and the return type \\
    Local Types* & types of local variables \\
    Numeric Literals* & numeric data constants used \\
    String Literals* & string data constants used \\
    Comments* & associated comment words \\
  }
\end{table}

\paragraph*{Type--Operation Coupling}
The feature-observation for this feature-class consists of the types
of variables operated on in the function, coupled with the operations
performed on those types.  The feature-observation is a set of (type,
operation) pairs.  Primitive types are paired with the built-in
arithmetic, logical, and relational operations, for example,
(\lstinline[keywordstyle=]{int}, \lstinline{>=}).  User-defined types
such as C++ classes are paired with the user-defined operations on
them, including direct and indirect field accesses and method calls.
For example, the pair (\lstinline{Bar}, \lstinline{.foo}) indicates
that the field \lstinline{foo} of an aggregate data type
\lstinline{Bar} is accessed.  The intuition behind including this
feature-class is that similar functions tend to use
similar type--operation pairs. 
For the example in \cref{fig:code-example}, the Type--Operation
Coupling feature-observation extracted is the set
\{(\lstinline[keywordstyle=]{int},~\lstinline{unary-}),
(\lstinline[keywordstyle=]{int},~\lstinline{/}),
(\lstinline[keywordstyle=]{int*},~\lstinline{+}),
(\lstinline[keywordstyle=]{int},~\lstinline{>}),
(\lstinline[keywordstyle=]{int},~\lstinline{+}),
(\lstinline[keywordstyle=]{int},~\lstinline{<=}),
(\lstinline[keywordstyle=]{int},~\lstinline{-}),
(\lstinline[keywordstyle=]{int},~\lstinline{<})\}.

\paragraph*{Skeleton Tree}
The feature-observation for this feature-class is based on the
abstract syntax tree (AST) of a function.  The AST is further
abstracted by retaining only the loops
(\lstinline[keywordstyle=]{for}, \lstinline[keywordstyle=]{while},
\lstinline[keywordstyle=]{do}\dots\lstinline[keywordstyle=]{while})
and conditionals
(\lstinline[keywordstyle=]{if}\ldots\lstinline[keywordstyle=]{else},
\lstinline[keywordstyle=]{switch}).  Operationally, the feature
extractor can be realized as a tree transducer that drops all AST
nodes that are not loops or conditionals.  Sequences of loops or
conditionals are encapsulated within a sequence node, and empty
sequences are dropped from the feature-observation.  The intuition
behind using this feature-class for code search is that similar
functions tend to have similar loop and conditional structures.

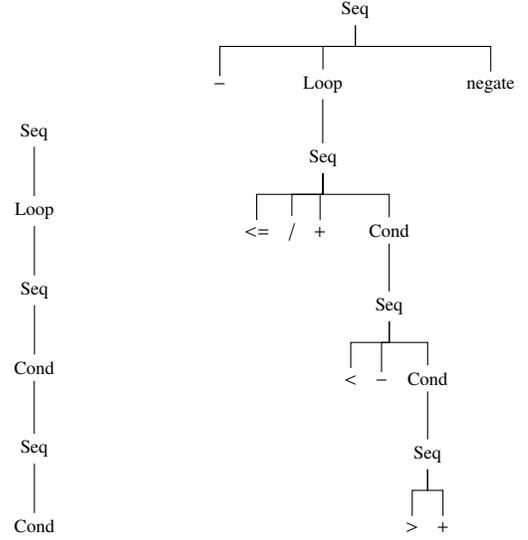
\begin{figure}[tb]
  \scriptsize
  \centering
  \hspace*{\fill}
  \subcaptionbox{Skeleton Tree\label{fig:tree-observations-undecorated}}[.8in]{%
    \begin{tikzpicture}
      \Tree [.Seq [.Loop [.Seq [.Cond [.Seq Cond ] ] ] ] ]
    \end{tikzpicture}
  }
  \hfill\hfill
  \subcaptionbox{Decorated Skeleton Tree\label{fig:tree-observations-decorated}}{%
    \begin{tikzpicture}[
      edge from parent/.style={
        draw,
        edge from parent path={
          (\tikzparentnode.south) -- +(0,-8pt) -| (\tikzchildnode)
        }
      },
      level distance=28pt,
      ]
      \Tree [.Seq {$-$} 
      [.Loop [.Seq {$<=$} {$/$} {$+$} [.Cond [.Seq {$<$} {$-$} [.Cond [.Seq {$>$} {$+$} ] ] ] ] ] ]
      negate ]
    \end{tikzpicture}
  }
  \hspace*{\fill}
  \caption{Tree-structured feature-observations for the example program in \cref{fig:code-example}.}
\end{figure}

\Cref{fig:tree-observations-undecorated} shows the Skeleton Tree feature-observation for the example code in \cref{fig:code-example}.

The similarity function used for Skeleton Tree feature-observations is based on tree edit distances.
Let $d_{\textit{r}}$ be a rough approximation of the distance between two trees, only based on their sizes:
\[
d_{\textit{r}}(T_1, T_2) = \frac{|\Size{T_1} - \Size{T_2}|}{\max(\Size{T_1}, \Size{T_2})}
\]
Further, let $D_T$ be a fixed distance threshold (which we set to $0.5$).
We obtain an approximate distance between two trees, $d_t$ as follows:
\[
  \scalebox{.96}{%
    \ensuremath{%
      d_{\emph{t}}(T_1, T_2) =
      \begin{cases}
        d_{\textit{r}}(T_1, T_2) & \mbox{if } d_{\textit{r}}(T_1, T_2) \geq D_{\textit{T}} \\
        \dfrac{\max
          \begin{pmatrix}
            \Ed{\Pre{T_1}}{\Pre{T_2}}, \\
            \Ed{\Post{T_1}}{\Post{T_2}}
          \end{pmatrix}
        }{\max(\Size{T_1}, \Size{T_2})}  & \mbox{otherwise}
      \end{cases}
    }
  }
\]
Here \Pre{T} is the sequence obtained by performing a pre-order traversal of the tree $T$, \Post{T} is the sequence obtained by performing a post-order traversal of the tree $T$, and \Ed{S_1}{S_2} is the word edit distance between the sequences $S_1$ and $S_2$.
The similarity function used for Skeleton Tree feature-observations is then computed as:
\begin{equation}\label{eq:sim-tree}
sim_{\emph{tree}}(T_1, T_2) = 1 - d_t(T_1, T_2)
\end{equation}

An exact tree-edit-distance computation~\cite{Zhang1989} has quartic-time complexity in the size of the trees being compared.  We instead
use a fast under-approximation of edit
distance~\citep{guha2002approximate} that gives our similarity
function quadratic-time complexity overall.  Note that we also use a
further rough approximation based on just the size of the trees, if
one of the two trees being compared is at least twice as large as the
other.  We found that using these approximations as opposed to the
exact tree-edit-distance based similarity made no discernible
difference in the quality of the final search results obtained, but
made a big difference in performance: more than $6\times$ faster in
our tests.

\paragraph*{Decorated Skeleton Tree}

This feature-class is similar to the Skeleton Tree, except that
instead of retaining just the loop and conditional structure in the
feature-observations, most operations (e.g., \lstinline{+},
\lstinline{-}, and \lstinline{<}) are also retained from the AST\@.
We discard some common operations, such as assignment
(\lstinline{=}) and address-of (\lstinline{&}), because they cause
excessive bloat.  The intuition behind including this feature-class 
is that similar functions use similar operations in
structurally similar locations.

\Cref{fig:tree-observations-decorated} shows the Decorated Skeleton
Tree feature-observation for the example code in
\cref{fig:code-example}.  The similarity function used is
$sim_{\emph{tree}}$ from \cref{eq:sim-tree}.

\paragraph*{Weighted NL Terms}
The feature-observations for this feature-class consist of various
natural-language (NL) terms in source code, such as function name,
comments, local variable names, and parameter names of a function.
Such NL terms, after extraction, are subjected to a series of standard
NL pre-processing steps, such as splitting words with under\_scores or
CamelCase, stemming, lemmatization, and removing single-character
strings and stop-words.  Stop-word removal discards both typical
English stop words such as ``the'', ``and'', and ``is''
\citep{NLTK-Stop-Words}, as well as stop words specialized for code,
such as ``fixme'', ``todo'', and ``xxx''.  Additionally, we use a
greedy algorithm~\cite{feild2006empirical} for splitting terms into
multiple words based on dictionary lookup.  This splitting is to
handle the case where programmers choose identifiers that combine
multiple words without under\_scores or CamelCase.

After NL pre-processing, we compute a \emph{term frequency-inverse
  document frequency} (TF-IDF) score for each NL term.  We consider
each function as a document, and compute the TF-IDF per C/C++ project.
We give function-name terms an inflated score ($5\times$ more than
other terms) because these often provide significant information about
functions' purposes.  The intuition behind including this
feature-class is that similar functions tend to have similar
natural-language vocabulary.  The feature-observation for the example
in \cref{fig:code-example} is \{``\lstinline{bin}'': $0.65$,
``\lstinline{search}'': $0.65$, ``\lstinline{high}'': $0.13$,
``\lstinline{low}'': $0.13$, ``\lstinline{found}'': $0.13$,
``\lstinline{mid}'': $0.13$, ``\lstinline{match}'': $0.13$\}.

The similarity function for two observations of Weighted NL Terms uses cosine similarity:
\[
sim_{\emph{nl}}(A, B) = \frac{\sum_{i=1}^{n} A_i B_i}{\sqrt{\sum_{i=1}^{n} A_i^2} \sqrt{\sum_{i=1}^{n} B_i^2}}
\]

Here $n$ is the total number of words in the universe, $A$ and $B$ are
vectors with TF-IDF scores, and the $i^{\text{th}}$ index $A_i$ is the
TF-IDF value for the $i^{\text{th}}$ word.

\paragraph*{K-Subgraphs of CFG} 

\begin{figure}[tb]
  \centering
  \hfill
  \begin{tikzpicture}[baseline=(B.center), level distance=8mm]
    \Tree [.A [ .\node(B){B}; [.C ] [.D ] ] ]
  \end{tikzpicture}
  \hfill
  \begin{tabular}{c|cccc} % chktex 44
    &   A & B & C & D \\ \hline % chktex 44
    A & 0 & 1 & 0 & 0 \\
    B & 0 & 0 & 1 & 1 \\
    C & 0 & 0 & 0 & 0 \\
    D & 0 & 0 & 0 & 0  
  \end{tabular}
  \hspace*{\fill}
  \caption{An example 4-graph and its corresponding adjacency matrix. Serializing the adjacency matrix entries yields binary digits ``0100 0011 0000 0000'', or \num{17152} in decimal. Node ordering in the adjacency matrix is the traversal order.\label{fig:graph-conversion}}
\end{figure}
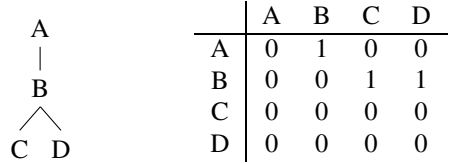

We implement multiple feature-classes based on $k$-sized subgraphs of the control flow graph (CFG) of a function.
Given the CFG of a function, we begin either a breadth-first-search (BFS) traversal or a depth-first search (DFS) traversal at a node until $k$ nodes are traversed; a subgraph of the CFG involving these $k$ nodes is extracted.
If fewer than $k$ nodes are reachable from a node (including itself), then such a sub-graph is thrown away.
We repeat this process for every node in the CFG, extracting at most $n$ subgraphs of size $k$, where $n$ is the size of the CFG\@.
We represent a graph of size $k$ as a $k^2$-bit integer, which is a 1-D  representation of a 2-D adjacency-matrix representation of the graph, obtained by concatenating each of the matrix rows in order.
Thus, from each function's CFG, we extract a multiset of $k$-graph shapes.
\Cref{fig:graph-conversion} shows an example of converting a 4-graph into a $16$-bit integer in this manner.

We implement the following four feature-classes based on the value of $k$ and the traversal strategy chosen: 
\begin{description}
  \item [\textbf{3 Graph CFG BFS}:] $k=3$, traversal strategy is BFS\@.
  \item [\textbf{4 Graph CFG BFS}:] $k=4$, traversal strategy is BFS\@.
  \item [\textbf{3 Graph CFG DFS}:] $k=3$, traversal strategy is DFS\@.
  \item [\textbf{4 Graph CFG DFS}:] $k=4$, traversal strategy is DFS\@.
\end{description}

For the example in \cref{fig:code-example}, the feature-observation
extracted for the feature-class ``4 Graph CFG BFS'' is the multiset
\{134, 134, 134, 194, 194, 194, 194, 194, 2114, 2114, 2114\}.  The
intuition behind including these feature-classes is that similar
functions tend to have similar control-flow structures
\citep{Khoo:2013:RSE:2487085.2487147}.

The similarity function used for these feature-class is based on the generalized Jaccard index between two multisets $O_1$ and $O_2$:
\begin{equation} \label{eq:generalized-jaccard}
sim_{\emph{Gen-Jacc}(O_1, O_2)} = \frac{\sum_{i} \min(O_{1i}, O_{2i})}{\sum_{i} \max(O_{1i}, O_{2i})}
\end{equation}
Here, $i$ iterates over all the unique elements in $O_1 \cup O_2$, and $O_{1i}$ is the number of times $i$ appeared in the multiset $O_1$.

\paragraph*{Calls to Library Functions}

We implement three feature-classes that extract calls to various kinds of library functions:
\begin{description}
\item [\textbf{Modeled Library Calls:}] CodeSonar models a large range of library functions for performing static analysis on C/C++ code. 
For this feature-class, calls made to any of these modeled library functions are extracted.  

\item [\textbf{Unmodeled Library Calls:}] Calls made to any unmodeled library functions are extracted for this 
feature-class---that is, calls to a function not modeled by CodeSonar, and whose definition is not available in the source code.

\item [\textbf{User-Defined Library Calls:}] For this feature-class, calls to functions whose definitions are available in a directory different from the caller function are extracted.
We use such functions as a heuristic for identifying user-defined libraries.
\end{description}

The intuition behind including the above three feature-classes is that
 similar code tends to call the same library functions.  For each
of these three feature-classes, the feature-values are sets of library functions
called.  A library function is represented as tuple: it includes the name
of the function together with the file name containing the
function's declaration.  For example, if a function calls
\lstinline{strcpy} and \lstinline{strncpy}, then the
feature-observation corresponding to Modeled Library Calls for that
function is \{(\lstinline{strcpy}, \lstinline{string.h}),
(\lstinline{strncpy}, \lstinline{string.h})\}.

\paragraph*{Type Signature}

For this feature-class, the feature-observations consist of the type
signature of the function: i.e., the argument types and the return
type of a function.  Together, the argument types and the return type
form a multiset of types.  For the example code in
\cref{fig:code-example}, the feature-observation corresponding to Type
Signatures is \{\lstinline[keywordstyle=]{int},
\lstinline[keywordstyle=]{int},
\lstinline[keywordstyle=]{int*}, \lstinline[keywordstyle=]{int}\}.
Type signatures define a function's interface for interaction with the
rest of the code.  Similar code tends to have similar interfaces, and
therefore type signatures could help with code search.

The generalized Jaccard index (\cref{eq:generalized-jaccard}) is used as the similarity function for this feature-class.

\paragraph*{Local Types}

For this feature-class, the feature-observations consist of the set of 
types of all the local variables. The intuition behind using local variable
types in code search is that similar code creates and operates on variables of 
similar types.  For the example code in \cref{fig:code-example}, the
Local Types feature-observation is \{\lstinline[keywordstyle=]{int}\}.

\paragraph*{Constants}
We implement two feature-classes that extract constants from a function: 
\begin{description}
\item [\textbf{Numeric Literals:}] This feature-class is described in \cref{Se:offline}.
\item [\textbf{String Literals:}] For this feature-class, a feature-observation is the set of all the literal strings used in a function. 
\end{description}
The intuition behind using sets of constants in code search is that
similar code typically uses similar constants.

\paragraph*{Comments}
For this feature-class, the feature-observations consist of the comments associated with a function.
The comments are represented as a set of words. 
The intuition behind using comments in code search is that the comments in similar pieces of code are likely to use a similar vocabulary.
For the example code in \cref{fig:code-example}, the Comments feature-observation is \{``found'', ``match'', ``no''\}.

\paragraph*{Combining Feature-Classes}

Using several feature-classes in combination allows \name{} to obtain
good code-search results in a fairly robust manner by using different
dimensions of the code.  For example, consider the binary-search
implementation in \cref{fig:code-example}.  We see that variables
named \lstinline{mid}, \lstinline{low}, \lstinline{high} are used;
that there are two conditionals nested inside a single loop; and that
an integer division and integer less-than-or-equal-to operation is
performed.  When put together, these observations are hallmarks of a
binary-search implementation.

\subsection{Dynamic Feature-Class Selection} 
\label{Se:PerQuerySelection}

Combining feature-classes can be beneficial for code search, however,
the feature-classes that are useful for performing a code search may
vary from one query to another.  For example, consider a query
function containing of just straight-line code.  A significant number
of functions in our code-database are devoid of loops and
conditionals,\footnote{We did a brief study of feature-observation
  distributions for the Skeleton Tree feature-class over our corpus,
  which revealed this data point.} and all such functions look
identical to the query function with respect to the Skeleton Tree
feature-class.  Thus, performing a code search with this query by
including the Skeleton Tree feature-class can lead to lower-quality
results.  On the other hand, if a query function has an unusual loop
and conditional structure that is idiomatic to the computation being
performed, then the Skeleton Tree feature-class would be useful in
code search: other instances of the same distinctive structure from
the code database would have high similarity scores to the query
function.

Thus, it is useful to select feature-classes automatically on a
per-query basis for code search.  This configuration of \name{} is
called \dynSelect{}.  Intuitively, a feature-class for a given
query is selected for code search if the corresponding
feature-observation is sufficiently discriminatory/unique with respect
to the overall feature-observation distribution for that
feature-class.

To prepare for the dynamic feature-class selection on a per-query basis, we take following steps offline:
\begin{itemize}
\item From the code database, we retrieve a random sample $S$ of
  feature-vectors. Random
  sampling gives an inexpensive estimate of feature-observation
  distributions across the entire code database.

\item We calculate a similarity threshold for each feature-class $c$
  by (1) computing pairwise similarity scores on the
  feature-observations for $c$ in $S$ and (2) taking the sum of means
  and standard deviations of the similarity scores.  Two
  feature-observations for $c$ are considered similar if their
  similarity score is above the similarity threshold for $c$.
\end{itemize}

Online, when a query is posed, we take the following steps for each feature-class $c$ (which can be performed in parallel):
\begin{itemize}
\item We compare the query's feature-observation for $c$ with all other feature-observations for $c$ in sample $S$ (of size $n_{\textit{samp}}$), and count the number of similar feature-observations $n_{\textit{sim-c}}$.
\item We select the feature-class $c$ for code search if it is not too
  common, that is, if
  $\frac{n_{\textit{sim-c}}}{n_{\textit{samp}}} < t_{\textit{uniq}}$.
  Here $t_\textit{uniq}$ is a threshold that indicates a
  feature-observation is sufficiently unique in the sample.  For
  example, $t_\textit{uniq} = 0.15$ indicates that any
  feature-observation that is similar to less than $15\%$ of the
  sample feature-observations is considered distinctive enough to
  warrant inclusion.
\end{itemize}

Each feature-class is assigned a weight of exactly $1.0$ or exactly
$0.0$, based on whether the feature-class is selected in the above
process.  These weights are used for combining feature-class
similarities for code search (\cref{eq:combined-sim}), and the
$k$-most-similar-function search is carried out between the query
function and the functions in the code database as described in
\cref{Se:online-search}, to obtain the $k$ functions most similar to the
query.

\subsection{SVM-Guided Feature-Class Weight Generation}
\label{Se:svm-weights}

Note that \dynSelect{} does not need any additional knowledge about
the query.  However, if we know ahead of time that a query belongs to
a specific domain, and we have ground-truth information available
regarding what constitutes similar code in that domain, then we can
use supervised-learning techniques to learn good feature-class weights
(for \cref{eq:combined-sim}) for that domain ahead of time, and use
these weights for code search with all future queries in that domain.

Given a particular ground-truth data set with labeled similar code, we
generate fine-tuned weights by training a binary-classification
support vector machine (SVM).  We do not train using raw code text, or
even raw sets of feature-observations.  Because we use the SVM
training process to generate relative weights for feature-class
similarity scores in \cref{eq:combined-sim}, we train the SVM on these
similarity scores directly.  The similarity scores for all
feature-classes between two functions are assembled into a
\emph{similarity vector}.  The SVM is then trained on examples of
similarity vectors for both similar and dissimilar functions, each
labeled accordingly.  This technique allows us to optimize ahead of
time how these feature-classes are relatively weighted in a code
search, by using the same similarity functions that are employed in
code search of a query.

Our SVM uses a linear classifier, which allows a convenient
interpretation of internal
weights~\citep{Guyon:2003:IVF:944919.944968}.  The final pre-processing step is
to extract these internal weights and normalize them relative to the
sum of their magnitudes, truncating negative weights.  These
normalized weights are then used directly as feature-class weights in \cref{eq:combined-sim}.

\Cref{Se:ExperimentalSetup} provides more details about the corpus and
training process. 
Of course, it is not obvious that weights obtained by training for 
classification purposes are useful in ranking results for
code-search queries. 
\Cref{Se:Results} measures 
the effectiveness of this strategy in practice.

%!TEX root = paper.tex
\section{Experimental Evaluation}
\label{Se:Experiments}

This section outlines the research questions we seek to answer through experiments
(\sectref{ResearchQuestions});
describes the setup and methodology used in the experiments
(\sectref{ExperimentalSetup}); and
presents the results of the experiments
(\sectref{Results}).

\begin{table}
  \centering
  \caption{Task categories used for code-search queries in \algo.
    ``\#Similar'' gives the number of similar functions that were manually found for a given task category.
    ``Partial'' reports how many function pairs MOSS considered to be potential clones.
    ``Significant'' reports the \% of function pairs with at least 50\% code overlap.\label{tab:tasks}}
  \begin{filecontents*}{raw_data/categories.csv}
Task Category,#Similar,Partial,Significant
Binary Search,5,.2,.1
Edit Distance,5,.4,0
Insertion Sort,5,.3,.3
Knapsack,5,.1,0
Modular Exponentiation,6,0,0
Non Recursive Depth First Search,5,0,0
Red Black Tree Left Rotate,6,.4,.13
\end{filecontents*}
\pgfplotstabletypeset[
  columns/Task Category/.style=text column,
  columns/Similar/.style={
    column name=\#Similar,
    column type={S[table-format=1]},
  },
  columns/Partial/.style=percentage column,
  columns/Significant/.style=percentage column,
  every head row/.append style={
    before row={%
      \toprule
      & & \multicolumn{2}{c}{MOSS Detected} \\
      \cmidrule{3-4}
    },
  },
  ignore chars=\#,
  ]{raw_data/categories.csv}
\end{table}

\subsection{Research Questions}
\label{Se:ResearchQuestions}

Our experiments were designed to answer the following research
questions:
\begin{enumerate}[label={\bfseries RQ\arabic*}]
\item\label{RQ:SoloClasses}\emph{How do the individual feature-classes
    described in \cref{Se:CodeSearch} perform in code-search tasks
    relative to each other?}

\item\label{RQ:CombinedClassesDynamic}\emph{Does combining
    feature-classes using per-query dynamic feature-class selection
    (\cref{Se:PerQuerySelection}) improve \name's performance?}

\item\label{RQ:SVMWeights}\emph{Does combining feature-classes using
    supervised learning (\cref{Se:svm-weights}) further improve
    \name's performance, when the query domain is known?}

\Omit{
\item\label{RQ:NoNL}\emph{How does \name{} perform in the total
    absence of natural language related feature-classes?}  \warn{We
    might do away with this research question for the lack of space.}
}
\end{enumerate}

A code-search task involves searching for
relevant documents from a group of documents that include both
relevant and non-relevant documents.
(In the case of \name, ``documents'' are C/C++ functions.)
Non-relevant documents are also known as \emph{distractors},
which leads naturally to the following question:
\begin{enumerate}[resume,label={\bfseries RQ\arabic*}]
\item\label{RQ:Distractors}\emph{How much does \name's code-search
    performance degrade as we increase the number of distractors in
    the code base being searched?}
\end{enumerate}

\subsection{Experimental Setup and Methodology}
\label{Se:ExperimentalSetup}

\name{} uses CodeSonar, an industrial-strength C/C++ static-analysis
engine, to analyze C/C++ corpora and implement feature extractors.
CodeSonar handles real-world C/C++ projects with tens of millions of
lines of code.  CodeSonar also exposes a wealth of information about a
program through well-defined APIs.  \name's feature extractors are
implemented as CodeSonar plugins that use these APIs.  Consequently,
\name{} inherits CodeSonar's requirement that programs must be
compilable to be analyzable.

\paragraph*{Code-Search Tasks}

Our experiments assess \name{}'s performance under various
configurations.  Code-search tasks are set up as follows.  
For each query function, there is a set of known relevant functions
that are similar to the query.
The relevant functions are treated as ground truth.
  The relevant functions are then mixed with
many non-relevant functions as distractors, and together they form the
code database used in the experiment.  \name{} then searches the code database for similar
functions.  We compute information-retrieval statistics based on the ranking of the known-relevant functions in the returned results. 

% This float is not used until much later.  We place it here in the
% LaTeX sources so that LaTeX will put it on the proper output page.
\Omit{
\begin{figure*}[tb]
  \centering
  \begin{tikzpicture}
    \begin{axis}[
      categorical retrieval plot,
      height=\axisdefaultheight,
      legend style={
        anchor=south east,
        at={(.97,.03)},
      },
      width=\linewidth,
      xticklabels={
        Insertion Sort,
        Edit Distance,
        Non Recursive\\Depth First Search,
        Knapsack,
        Binary Search,
        Modular Exp,
        RB Tree\\Left Rotate,
        Mean
      },
      ]
      \pgfplotstableread{raw_data/weight_impact.csv}{\table}
      \foreach \series in {R-Precision-Auto,R-Precision-Static,MAP-Auto,MAP-Static} { 
        \addplot+ table [y=\series] {\table} ;
        \edef\temp{\noexpand\addlegendentry{\series}}\temp
      }
    \end{axis}
  \end{tikzpicture}
  \caption{Retrieval statistics for each of the task categories from \cref{tab:tasks},
    showing the performance of \name{} configurations that use Auto-weighting 
    vs.\ Static-weighting.
    The size of the distractor set is \num{10000} functions.\label{fig:weight-impact}}
\end{figure*}
}

\paragraph*{Queries}
We use two query ground-truth sets for the code-search tasks,
representing two domains.  One, called \algo{}, represents
``algorithmic'' code queries.  For \algo{}, we created seven tasks,
outlined in \cref{tab:tasks}, and manually curated a total of
thirty-eight functions that each accomplish one of the seven tasks.  The
functions were mostly obtained from
\href{http://www.github.com/}{GitHub}, and were written by a variety of
programmers, none of whom are authors of this paper.  The functions
that accomplish a specific task have been manually vetted to be
similar to each other.  We thus have a total of thirty-eight base
queries.

We use these sets of real-world functions as queries (and 
the desired search results), and consider them to be an appropriate proxy for the code-search queries performed (and search results expected) by users in the algorithm domain.
We have made the labeled queries available for inspection.\footnote{Available at the URL\@: \url{http://tinyurl.com/source-forager-algo-benchmarks}.}

To make sure that the similar functions we found were not all clones
of each other, we ran them through the MOSS software-plagiarism
detector \citep{SIGMOD:SWA2003}.  Given a group of programs, MOSS
reports program pairs that may be clones, along with an overlap
percentage.  \Cref{tab:tasks} reports MOSS's findings, run using
default settings.  In this table, \emph{partial overlap} represents
any pair that MOSS reports as possible clones, while \emph{significant
  overlap} counts only possible clones with at least $50\%$ overlap.
Observe that many function pairs marked manually as being similar are
\emph{not} just MOSS-detectable clones of each other.  Thus,
recognizing similar function pairs in this corpus is a nontrivial
challenge.

The second query ground-truth set we use is called \libc{}, and
represents code queries from systems programming.  We looked at three
implementations of the standard C library: musl libc~\cite{musl-libc},
diet libc~\cite{diet-libc}, and uClibc~\cite{uclibc}.  From these we
define $88$ function categories corresponding to $88$ functions that
all three implementations provide.  We assume that within the same
function category, the three libc implementations are ``similar.''
For this domain, we have $88 \times 3$ queries.  For example, musl
libc's \lstinline{sprintf} is labeled to be similar to diet libc's \lstinline{sprintf} and uClibc's \lstinline{sprintf}, and dissimilar to everything else.

\paragraph*{Distractor Functions}
The distractor functions have been taken from the openly available
MUSE corpus~\citep{MUSE-Corpus}, and mainly consist of code from
Fedora source packages (SRPMs).  Our feature extractors currently
require compilable code, which Fedora SRPMs provide.  Due to the large size of the
distractor-function corpus (over \num{200000}), we have not
manually vetted all of the distractor functions to be sure that they
are irrelevant to the queries issued.  It is possible that
some distractor functions are indeed relevant to some queries, so our
retrieval statistics are under-approximations.
With the exception of the experiments reported in \cref{fig:noise-impact}, all experiments use \num{10000} distractors.

\paragraph*{Retrieval Statistics}
We compute Mean Average Precision (MAP) as the retrieval statistic, as
is common in information retrieval.  MAP is typically used to measure
the quality of \emph{ranked} retrieval results, because MAP takes into
account the rank of the relevant documents in the retrieved results.
MAP provides a measure of quality across all recall levels.  MAP is
the mean of the average precision computed for each query.  The
average precision (AP) for each query is given by
$\left(\sum_{k=1}^{n} P(k) \cdot r(k)\right) / R$, where $n$ is the
total number of documents searched; $R$ is the number of documents
marked relevant to the query; $P(k)$ is the precision when $k$
documents are requested; and $r(k)$ is $1$ when the $k^\text{th}$
retrieved document is relevant, and $0$ otherwise.  That is, AP is the
average precision at all the points when a new relevant document is
retrieved in a ranked result list.
%Relevant documents that are ranked poorly, i.e., those that contribute towards poor recall, 
%also contribute towards lowering the average precision.
The best MAP score that can be
achieved is $1.0$, when for each query, the $R$ relevant documents
appear as the top $R$ search results.

\Omit{
\begin{description}
  \item [R-Precision:]
    This statistic computes the precision (number of relevant retrieved
    documents divided by the number of retrieved documents) of a retrieval
    task, when the number of documents requested is set to the actual
    number of relevant documents, $R$.
    The higher the R-Precision value, the better the quality of the
    search results.
    The highest possible R-Precision value is $1.0$, when precisely
    the $R$ relevant documents are retrieved, with no omissions and no
    irrelevant extras.
    The R-Precision is also the point in the precision-recall
    trade-off when the precision, recall (ratio of number of relevant
    retrieved documents to the number of relevant documents), and F1-score
    (which is the harmonic mean of precision and recall) are all the same.
    The R-Precision score for a set of queries is the average of the
    R-Precision for the individual queries.
    This measure is known to be typically correlated with MAP, defined
    next.

  \item [Mean Average Precision (MAP):]
  
  \item [Recall--Precision Graph:]
    This graph maps the interpolated precision at different recall
    levels ranging from $0.0$ to $1.0$, in increments of $0.1$.
    The interpolated precision at a recall level $r$ is defined as the
    highest precision found for any recall level $r' \geq r$.
    A graph is obtained over multiple queries by averaging the precision
    across same recall levels.
\end{description}
}

\paragraph*{SVM-Guided Weights}

We applied the techniques discussed in \cref{Se:svm-weights} on \algo{} and \libc{} to provide labeled data-sets on which to
train an SVM\@.  
Each instance in our training set is generated by
comparing two functions $a$ and $b$, yielding a single similarity
vector that consists of similarity scores for each feature-class.  The
binary classification for each training instance is $1.0$ if $a$ and
$b$ are implementations of the same function, $0.0$ otherwise.

We use LIBLINEAR~\citep{REF08a}
to train the SVM to classify these function comparisons;   
this process takes roughly twenty milliseconds.  Using 
this technique, we are able to achieve over 98\% accuracy under
ten-fold cross-validation.

Once the SVM is trained, we extract and normalize its internal weights
for use in code search.  For the \svmWeights{} configuration described
below, within each domain, the data-set is divided into multiple folds
of training-set and test-set pairs.  The weights extracted from the
training set are used to obtain MAP scores on the test set.  That is,
weights are trained on a subset of a given domain (\algo{} or \libc{})
and tested using queries from a different subset of the same domain.
For the \svmWeightsCross{} configuration described below, \algo{} is
used to train weights for queries from \libc{}, and vice-versa.
\paragraph*{\name{} Configurations}
\label{subsec:configurations}

Our experiments run \name{} under many configurations.  Each
configuration is defined by the weight $w_c$ assigned to each of the
feature-classes $c$ given in \cref{tab:feature-classes}.  These
weights are used in \cref{eq:combined-sim} for performing the
code search.

\begin{description}
  \item [\soloC{}: ] For each query, the weight $w_c$ corresponding to feature-class $c$ is $1.0$.  Weights corresponding to all other feature-classes are set to $0.0$.
  \item [\equalAll{}:] For each query, for all feature-classes $c$, $w_c = 1.0$, giving equal importance to all feature-classes for all queries.
  \item [\dynSelect{}:] For each query, a subset of feature-classes
    are selected and given equal weights, as described in
    \cref{Se:PerQuerySelection}.  The dynamic selection of
    feature-classes adds a small run-time overhead to each
    query.\footnote{In our naive Python implementation, \dynSelect{} adds an average
      run-time overhead of $2.1$ seconds per query for dynamic
      selection of feature-classes. Currently, the
      selection decision on each feature-class is done sequentially,
      instead of in parallel as suggested in \cref{Se:PerQuerySelection}.}
  \item [\randSelect{}:] For each query, a new random configuration is used as follows: a random subset of the feature-classes is selected, and the selected feature-classes are given equal weights. Repeat this process $10$ times with different random selections, and report mean results over these $10$ trials.
  \item [\svmWeights{}:] For each query, use weights learned for the domain that the query belongs to, as described in \cref{Se:svm-weights} and above.
  \item [\svmWeightsCross{}:] For each query, use weights learned for the domain that the query does \emph{not} belong to.

\end{description}
Note that, unlike the other configurations, the \svmWeights{} and \svmWeightsCross{} configurations permit weights to give different (non-zero) importance levels to different feature-classes.

\subsection{Results and Discussion}
\label{Se:Results}

%\pgfplotstabletypeset[columns/Feature-Classes/.style=string type]{\table}

\begin{figure*}[tb]
  \centering
  \begin{tikzpicture}
    \pgfplotstableread{raw_data/map-scores-10k-distractors.csv}{\table}
    \begin{axis}[
      ybar=0,
      retrieval plot,
      bar width=8pt,
      width=\linewidth,
      height=\axisdefaultheight,
      table/x expr={\coordindex + (\coordindex >= 16) / 2},
      xmin=-.75,
      xmax=21.25,
      minor xtick=15.75,
      xminorgrids,
      minor x grid style=black,
      x tick label style={
        align=right,
        font=\scriptsize,
        text width=.63in,
      },
      xticklabel={%
        \pgfplotstablegetelem{\ticknum}{feature-class}\of\table
        \ifdefempty{\pgfplotsretval}{
          \pgfplotstablegetelem{\ticknum}{configuration}\of\table
          \configuration{\pgfplotsretval}
        }{\pgfplotsretval}
      },
      haloed nodes near coords,
      every node near coord/.append style={anchor=west},
      legend style={
        anchor=north,
        at={(.5,.97)},
      },
      ]
      \addplot+ table[y=algo-qs]{\table} ; \addlegendentry{\algo}
      \addplot+ table[y=libc-qs]{\table} ; \addlegendentry{\libc}
    \end{axis}
  \end{tikzpicture}
  \caption{Information retrieval performance with \num{10000}
    distractors.  The left side of the plot, from ``Type--Operation
    Coupling'' through ``Comments'', uses the \soloC{} configuration
    with the given feature-class as the only non-zero-weighted feature.
    The right side of the plot, from ``\equalAll{}'' through
    ``\svmWeightsCross{}'', uses the various other \name{}
    configurations that leverage multiple feature-classes
    simultaneously.  Although no MAP score is exactly zero, several are
    below $0.005$ and therefore round to ``$0.00$'' in the data-point
    labels above.\label{fig:map-scores-10k-distractors}}
\end{figure*}

The left side of \cref{fig:map-scores-10k-distractors} shows how each
individual feature-class performs on the code-search tasks in isolation.
This experiment addresses \cref{RQ:SoloClasses}.  The solo feature-class
Weighted NL Terms ($0.70$, $0.86$)\footnote{Pairs of numbers following a
  configuration in this section indicates the MAP scores of that
  configuration on \algo{} and \libc{}, respectively.}  performs the best
individually on both \algo{} and \libc{}.  Thus:

\begin{finding}{RQ:SoloClasses}
  If we were to drive \name{} using only one feature-class, Weighted
  NL Terms is the best option.  However,
  \cref{fig:map-scores-10k-distractors} shows that the performance of
  the different feature-classes varies considerably depending on the
  query ground-truth set.  This variance suggests that different
  feature-classes are important for different kinds of queries.
\end{finding}

\Omit{
\begin{figure}[tb]
  \centering

  \begin{tikzpicture}
    \begin{axis}[
      numeric retrieval plot,
      every axis plot/.append style={
        fill,
        fill opacity=.1,
        mark options={opacity=1},
      },
      x tick label style={rotate=-90},
      xlabel=Recall,
      ylabel=Precision,
      ]
      \addplot+ [every node near coord/.append style={anchor=west}] table {raw_data/pr_curve.csv} \closedcycle ;
      \addlegendentry{Auto-Weighting}
      \addplot+ table [y=Static-Weighting] {raw_data/pr_curve.csv} \closedcycle ;
      \addlegendentry{Static-Weighting}
    \end{axis}
  \end{tikzpicture}
  \caption{Interpolated precision vs.\ recall levels for all of the \algo{} and \libc{} queries using the configurations \warn{which configurations to use here? Possibly \equalAll{}, \dynSelect{}, \svmWeights{}}.
    The greater the area under the curve, the better is the overall quality of the
    retrieval system under a range of recall levels. The size of the distractor set is \num{10000} functions.\label{fig:pr-curve}}
\end{figure}
}

\Cref{RQ:CombinedClassesDynamic} asks whether multiple feature-classes
can be usefully combined, and whether \dynSelect{} is a good way to do
such a combination.  A straight-forward manner in which
feature-classes can be combined is the \equalAll{} configuration,
which represents a baseline to compare against other configurations.
The \dynSelect{} configuration selects different subsets of the
feature-classes on a per-query basis (\cref{Se:PerQuerySelection}).
As a sanity check for the selections performed by \dynSelect{}, we
also compare it with the \randSelect{} configuration, which randomly
selects feature-class subsets for every query.  The right side of
\cref{fig:map-scores-10k-distractors} shows that \dynSelect{} ($0.84$,
$0.89$) performs better on both \algo{} and \libc{} when compared to
\equalAll{} ($0.67$, $0.73$) and \randSelect{} ($0.57$, $0.63$).
\dynSelect{} also outperforms each of the solo configurations from the
left side of \cref{fig:map-scores-10k-distractors}.

\begin{finding}{RQ:CombinedClassesDynamic}
  In the absence of any additional information about the query,
  combining multiple feature-classes and dynamically selecting
  feature-classes on a per-query basis (\cref{Se:PerQuerySelection})
  is the most effective strategy for code search.
\end{finding}

\Cref{RQ:SVMWeights} addresses the scenario where the domain of a
query is known, and additional information is available regarding that
domain (as described in \cref{Se:svm-weights}).  The \svmWeights{}
configuration tests \name{} under this scenario.  Pre-learning the
relative importance of feature-classes for a given domain (in the form
of weights $w_c$ for each feature-class) also makes code search more
efficient by eliminating any run-time overhead in feature-class
selection.  The right side of \cref{fig:map-scores-10k-distractors}
shows that \svmWeights{} ($0.86$, $0.95$) outperforms all other
configurations.

\Omit{
% we have already described the ideas below before
: we divide each of the \algo{} and
\libc{} data sets into multiple folds of training and test queries.
Relative weights are learned on the feature-classes using the training
set belonging to the same domain (using techniques outlined in
\cref{Se:svm-weights}), and the learned weights are used in reporting
scores over the test sets.
}

\begin{finding}{RQ:SVMWeights}
  When the domain of a query is known, and training data is available,
  combining multiple feature-classes using weights derived from
  supervised learning (\cref{Se:svm-weights}) is the most effective
  strategy for code search.
\end{finding}

The \svmWeightsCross{} ($0.74$, $0.85$) configuration tests whether the
weights learned from one domain are useful in a different domain.
\Omit{ This configuration uses the \libc{} data set to train weights that
  are employed to test \algo{}, and vice versa.  } The rightmost two
bars in \cref{fig:map-scores-10k-distractors} show that it is hard to
derive a single set of relative feature-class weights that work well for
queries in both domains. Thus, in the absence of domain information about
the query, \dynSelect{} is preferred.

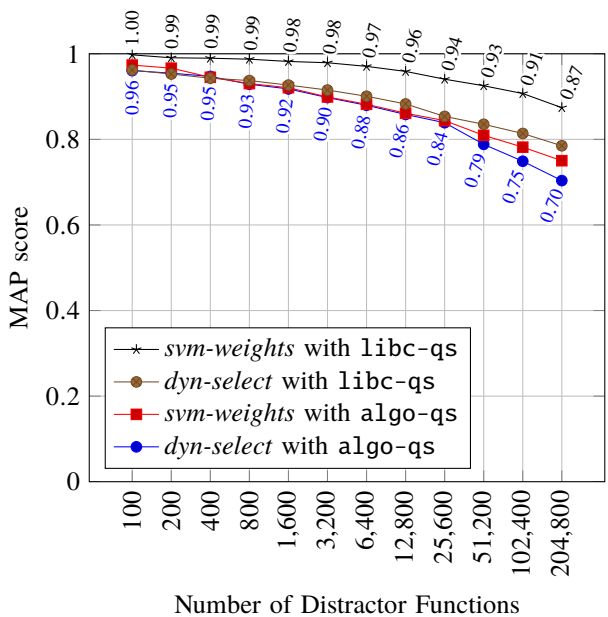
\begin{figure}[tb]
  \centering
  \begin{tikzpicture}
    \pgfplotstableread{raw_data/noise_impact.csv}{\table}
    \begin{axis}[
      legend style={
        anchor=south west,
        at={(.03,.03)},
      },
      retrieval plot,
      every node near coord/.append style={
        rotate={-2*\coordindex},
      },
      reverse legend,
      table/x=Distractor Corpus Size,
      table/x expr=\coordindex,
      xlabel=Number of Distractor Functions,
      xmajorgrids,
      xticklabels={100,200,400,800,{1,600},{3,200},{6,400},{12,800},{25,600},{51,200},{102,400},{204,800}},
%      x tick label style={rotate=-45, anchor=north east, inner sep=0mm},
      ]
      \addplot+ [haloed nodes near coords] table [y=dyn-select-algo] {\table} ;
      \addlegendentry{\dynSelect{} with \algo{}}
      \addplot+ table [y=svm-weights-algo] {\table} ;
      \addlegendentry{\svmWeights{} with \algo{}}
      \addplot+ table [y=dyn-select-libc] {\table} ;
      \addlegendentry{\dynSelect{} with \libc{}}
      \addplot+ [haloed nodes near coords, every node near coord/.append style={anchor=west}] table [y=svm-weights-libc] {\table} ;
      \addlegendentry{\svmWeights{} with \libc{}}
    \end{axis}
  \end{tikzpicture}
  \caption{Impact of the number of distractor functions on MAP scores
    using \dynSelect{} and \svmWeights{} for all \algo{} and \libc{}
    queries.  The horizontal axis is on a log
    scale.\label{fig:noise-impact}}
\end{figure}

\Cref{fig:noise-impact} shows how \name{}'s result quality scales with
increasing distractor-set sizes.  This experiment addresses
\cref{RQ:Distractors}.  \name{} is used in the \dynSelect{} and
\svmWeights{} configurations for this experiment.  As one would
expect, MAP scores decline as distractors proliferate.  However,
consider that relevant sets contain just $2$ to $6$ items competing
against distractor sets that are up to five orders of magnitude
larger.

\begin{finding}{RQ:Distractors}
  Resilient MAP scores indicate that \name{} returns high-quality
  results even when distractors outnumber relevant items by several
  orders of magnitude.
\end{finding}

\subsection{Threats to Validity}
\label{Se:threats}

The issue of whether evaluation benchmarks are appropriate is a
potential threat to the validity of any information retrieval system.
We mitigate this threat for \name{} in several ways.  First, we use
benchmark queries from two different domains, \algo{} and \libc{}.
Second, we use the MOSS plagiarism detector to show that our manually
labeled set of relevant functions in \algo{} are not trivial clones of
each other.  Third, we draw the \algo{} and \libc{} data sets from
real-world code written by arbitrary programmers, not artificial
programs written by us.

Feature-classes can be combined in various ways to perform code
searches.  We have explored part of the vast space of all such
combinations, and our results speak only to those we have tried.  We
find that the MAP scores of the configuration \dynSelect{} on both
\algo{} and \libc{} are good.  We designed the experiments with
\equalAll{} and \randSelect{} configurations to test whether the
selections made by \dynSelect{} are indeed necessary and useful, and
find that they are.

\Omit{
In this section, we discuss the limitations of the current incarnation of \name{}
(\sectref{Limitations}).
\name{} performs code search and code retrieval at the function level.
It would be useful to extend \name{} to operate on other code
elements, such as fragments of functions.
\name{} also currently requires that the query code be compilable;
this limitation could be overcome by building feature extractors using
fuzzy parsers~\citep{Joern}.
}
%%% Local Variables: 
%%% mode: latex
%%% End: 

%  LocalWords:  ResearchQuestions ExperimentalSetup RQ1 's url MOSS's
%  LocalWords:  PerQueryWeights SRPMs RQ F1 PDB 500k SoloClasses NL
%  LocalWords:  CombinedClasses nl CodeSearch PerQuerySelection 's tb

%!TEX root = paper.tex

\section{Related Work}
\label{Se:RelatedWork}

\paragraph*{Code-search engines} 
Several popular text-based code-search tools ``grep'' over tokenized
source code: GitHub, SearchCode, Open HUB, etc.  While these tools are
useful, they fall short in many use cases, as they do not exploit the
rich semantics of code.  For example, the top search results for the
term ``dfs'' on C code projects in GitHub yields function
declarations, macro names, and \lstinline[keywordstyle=]{#include}
directives that mention ``dfs'', but that are not actually useful.

The Sourcerer code-search engine~\citep{DMKD:LBN+2009} combines text-based search techniques with information about relations among programming ``entities'' like packages, classes, methods, and fields.
Sourcerer also uses fingerprints that capture some light-weight structural information about the code, such as depth of loop nesting and presence or absence of certain language constructs.
Queries in Sourcerer are text-based and are powered by Lucene (\url{http://lucene.apache.org}), as opposed to the code-based search by \name{}. 

Strathcona~\citep{ICSE:HM2005} returns relevant Java code examples to
developers learning to use complex object-oriented frameworks.  It
uses several heuristics based on class-inheritance hierarchies, method
calls, and type uses.  \name{} could also use the applicable
heuristics from Sourcerer and Strathcona as feature-classes, but
additionally demonstrates how to search using more complex structures,
such as decorated skeleton trees and CFG-subgraphs.

CodeGenie~\cite{ASE:LBO+2007,IST:LBO+2011} proposes test-driven code search, in which the user supplies a set of unit tests for the code component they want to find.
CodeGenie leverages Sourcerer~\citep{DMKD:LBN+2009} to perform keyword-based search; test cases refine these results.
\name{} could be used as a replacement for Sourcerer to perform similar code search in CodeGenie.

\begin{NoHyper}\Citeauthor{SEM:SEB2014}\end{NoHyper}~\cite{SEM:SEB2014,ASE:KSGB2015}
perform code search based on logical characterizations of programs'
I/O behaviors, obtained via symbolic execution.  A query consists of
concrete I/O pairs for the desired code fragment.  While this approach
precisely captures the semantics of the corpus elements, it does not
immediately handle some common programming constructs, such as loops
and global variables.  It also restricts the size of the program
elements in the corpus, because symbolic execution of larger elements
may lead to path explosion.  \name{} can easily be extended to use I/O
pairs as an additional feature-class in scenarios where the above
restrictions are acceptable.

XSnippet~\cite{OOPSLA:SC2006} and ParseWeb~\cite{ASE:TX2007} are specialized code-search engines: XSnippet looks specifically for code that
instantiates objects of given type in a given context, ParseWeb has a similar focus on code sequences that instantiate objects.
Codify~\cite{HCIIR:B2007} extracts and stores a large amount of metadata for each symbol in a program, and provides a user interface for querying that metadata.
Codify aids in understanding and browsing code.
The goal of \name{}'s code search is different from the above, i.e., to find source code similar to a query. 

\paragraph*{Code-clone detection}
\name's code searches differ from the typical clone detection problem
in that we are interested in finding code that has both semantic and
syntactic similarity.  Therefore, we use a range of feature-classes
that span from syntactic to semantic.  \name{}'s notion of similarity
does not neatly fall into any of the definitions of standard clone
types 1--4~\cite{Roy2009}.

\paragraph*{Similar-machine-code search}
Finding similar machine
code~\cite{Khoo:2013:RSE:2487085.2487147,David2014,eschweiler2016discovre,pewny2015cross}
is useful in finding known vulnerabilities in third-party code for
which source code is not available.  The primary difference in code
search at the source-level and machine-level is that machine code has
poorer syntactic, semantic, and structural information available
compared to source code.  As a result, while there is some overlap
between techniques, research on machine-code search is focused on
tackling different problems, such as how to do similar-machine-code
search across different CPU architectures, compiler optimizations,
compilers, operating systems, etc.

\paragraph*{SVM-based code-classification}
\citet{Rosenblum:2011:WTC:2041225.2041239} train SVMs with features
extracted from source code in the attempt to classify programs by
author.  \name{} builds on this idea by training an SVM with similarity 
scores derived from feature-observations, and then extracting internal weights from
the trained SVM to strengthen the combined similarity function used for code search.

%%% Local Variables: 
%%% mode: latex
%%% End: 

%  LocalWords:  Sourcerer Lucene Stollee et al Reiss XSnippet pre dfs
%  LocalWords:  CodeGenie ParseWeb Strathcona SearchCode 's

\Omit{
\Omit{
\section{Conclusion}
\label{Se:Conclusion}

\warn{Page limit is 10 pages + 2 for references.}
}
% Incorporate richer type information
% Program elements other than just functions
% Incorporate user feedback

%%% Local Variables: 
%%% mode: latex
%%% End: 

}

% push the references over to the new page


% Generated by IEEEtranN.bst, version: 1.13 (2008/09/30)

% ACM needs 'a single self-contained file'!
\end{document}